\documentclass[journal]{IEEEtran}
\usepackage[utf8]{inputenc}
\usepackage{multirow}
\usepackage{amsmath,amsfonts,xcolor,graphicx,hyperref,bm, adjustbox, tabularx}
\usepackage[square,sort&compress,comma,numbers]{natbib}
\usepackage{gensymb}
\usepackage[font=footnotesize, singlelinecheck=false]{caption}
\usepackage{float}
\usepackage{dblfloatfix}
\usepackage{xcolor}
\usepackage{xargs}  
\usepackage[colorinlistoftodos,prependcaption,textsize=tiny]{todonotes}

\usepackage[]{color-edits}
\addauthor{mh}{blue!50!red}
\addauthor{ah}{green!25!red}
\addauthor{ma}{red}

\title{Dense Recurrent Neural Networks for Accelerated MRI: History-Cognizant Unrolling of Optimization Algorithms}

\author{Seyed Amir Hossein Hosseini,~\IEEEmembership{Student Member,~IEEE,}
        Burhaneddin Yaman,~\IEEEmembership{Student Member,~IEEE,}\\
        Steen Moeller,
        Mingyi Hong,~\IEEEmembership{Member,~IEEE}
        and Mehmet Ak\c{c}akaya,~\IEEEmembership{Member,~IEEE}
        \vspace{-0.4cm}

\thanks{S. A. H. Hosseini, B. Yaman and M. Ak\c{c}akaya are with the Department of Electrical and Computer Engineering, and Center for Magnetic Resonance Research, University of Minnesota, Minneapolis, MN, 55455.
S. Moeller is with the Center for Magnetic Resonance Research, University of Minnesota, Minneapolis, MN, 55455.
M. Hong is with the Department of Electrical and Computer Engineering, University of Minnesota, Minneapolis, MN, 55455.
E-mails: \{hosse049, yaman013, moell018, mhong, akcakaya\}@umn.edu}.
\thanks{This work was supported by NIH P41EB027061, NIH U01EB025144, NSF CAREER CCF-1651825, NSF CMMI-1727757, CIF-1910385, ARO grant 73202-CS.}}

\begin{document}
\graphicspath{{Figures/}}

\maketitle

\begin{abstract}
Inverse problems for accelerated MRI typically incorporate domain-specific knowledge about the forward encoding operator in a regularized reconstruction framework. 
Recently physics-driven deep learning (DL) methods have been proposed to use neural networks for data-driven regularization. These methods unroll iterative optimization algorithms to solve the inverse problem objective function, by alternating between domain-specific data consistency and data-driven regularization via neural networks. The whole unrolled network is then trained end-to-end to learn the parameters of the network. 
Due to simplicity of data consistency updates with gradient descent steps, proximal gradient descent (PGD) is a common approach to unroll physics-driven DL reconstruction methods. However, PGD methods have slow convergence rates,
necessitating a higher number of unrolled iterations, leading to memory issues in training and slower reconstruction times in testing. Inspired by efficient variants of PGD methods that use a history of the previous iterates, we propose a history-cognizant unrolling of the optimization algorithm with dense connections across iterations for improved performance. In our approach, the gradient descent steps are calculated at a trainable combination of the outputs of all the previous regularization units. We also apply this idea to unrolling variable splitting methods with quadratic relaxation. Our results in reconstruction of the fastMRI knee dataset show that the proposed history-cognizant approach 
reduces residual aliasing artifacts compared to its conventional unrolled counterpart without requiring extra computational power or increasing reconstruction time. 
\end{abstract}

\section{Introduction}
Inverse problems have been extensively utilized in many imaging modalities, including magnetic resonance imaging (MRI) \cite{pruessmann2001advances, sutton2003fast, lustig2007sparse,  block2007undersampled, akccakaya2010compressed,knoll2011second,akccakaya2011low}. These inverse problems are directly driven from the physics of data acquisition, known as the forward operator, which incorporates domain-specific knowledge. Such inverse problems are typically ill-conditioned and thus regularizers are incorporated into the respective objective functions. Subsequently, iterative algorithms are employed to solve the regularized inverse problem by alternating between enforcing physics-driven data consistency and regularization using pre-selected priors.

Recently, deep learning (DL) has emerged as an alternative state-of-the-art technique for solving such inverse problems in imaging applications \cite{wang2016accelerating,kwon2017parallel, lee2018deep,  wang2018image,dar2018transfer,chen2017low, han2019k,han2018deep, yang2018low, kang2018deep, wu2019simba, hammernik2018learning, schlemper2017deep, mardani2017recurrent, IstaNet, kellman2019physics, yang2018admm, deepADMMnet,ye2018deep, aggarwal2018modl, qin2018convolutional, duan2019vs, yaman2019self, yaman2019selfMRM, pdCTbyOktem, cheng2019model, bostan2018learning, kamilov2016learning, akcakaya2019scan, hosseini2020accelerated, hosseini2019sraki, knoll2019deep}. Several of the earlier works concentrated on training convolutional neural networks (CNN) to solve the problem without incorporating explicit knowledge of the acquisition physics, typically requiring re-training when acquisition parameters are modified
\cite{wang2016accelerating,kwon2017parallel, lee2018deep, dar2018transfer, wang2018image, chen2017low, han2019k, han2018deep, yang2018low, kang2018deep,  wu2019simba}. Another line of work focuses on a physics-driven formulation, where the known forward model is incorporated into training, utilizing domain knowledge to solve the corresponding inverse problem \cite{schlemper2017deep, qin2018convolutional, hammernik2018learning,  aggarwal2018modl, mardani2017recurrent, IstaNet,    yaman2019self, yaman2019selfMRM, yang2018admm, deepADMMnet, cheng2019model, duan2019vs, ye2018deep, pdCTbyOktem, kellman2019physics, bostan2018learning, kamilov2016learning}. In physics-driven approaches, a conventional iterative optimization algorithm for solving a regularized least-squares problem is unrolled for a fixed number of iterations. This unrolled network alternates between data consistency and regularization steps similar to the conventional algorithm, where the regularizers are implemented via neural networks. The unrolled network is trained end-to-end to learn the network parameters, which predominantly characterize the regularization units.

There are numerous approaches for solving regularized least squares problems, which have been explored for algorithm unrolling in physics-driven DL reconstruction methods \cite{schlemper2017deep, qin2018convolutional, hammernik2018learning, aggarwal2018modl, mardani2017recurrent,yaman2019self, yaman2019selfMRM,   IstaNet, yang2018admm, deepADMMnet, cheng2019model, duan2019vs, pdCTbyOktem, kellman2019physics, ye2018deep, bostan2018learning, kamilov2016learning}. These methods include gradient descent (GD) \cite{hammernik2018learning}, proximal gradient descent (PGD) \cite{mardani2017recurrent, IstaNet, kellman2019physics, schlemper2017deep}, variable-splitting (VS) \cite{yang2018admm, deepADMMnet,ye2018deep, aggarwal2018modl, qin2018convolutional, duan2019vs, yaman2019self, yaman2019selfMRM} and primal-dual (PD) \cite{pdCTbyOktem, cheng2019model} methods. The GD, PGD and PD unrolling typically use a gradient descent step for data consistency, while for VS-based methods, either a gradient descent \cite{ye2018deep} or conjugate gradient (CG) data consistency can be used, with the latter being utilized when the least squares problem involving the forward encoding operator becomes computationally expensive \cite{yang2018admm, deepADMMnet, aggarwal2018modl, qin2018convolutional, duan2019vs, yaman2019self, yaman2019selfMRM}. Data consistency updates with gradient descent step is generally computationally inexpensive. However, these optimization methods have slower convergence, which necessitates higher number of unrolled iterations for improved performance, which in turn can lead to numerical issues such as gradient vanishing and memory constraints. 

In this study, we seek to develop a new methodology for unrolling inverse problem optimization algorithms into dense recurrent neural networks (RNN) for improved reconstruction performance. Motivated by accelerated PGD approaches, which calculate the gradient descent at a combined history of previous iterations rather than the immediate previous iteration only \cite{nesterov2005smooth, nesterov2013gradient, nesterov2019gradient, beck2009fast,  Chambolle15, hong13complexity, tseng08acc, becker2011nesta}, we propose a history-cognizant unrolling of the optimization algorithms with skip connections across unrolled iterations, leading to a dense RNN architecture. Skip connections have been previously shown to facilitate information flow through neural networks in architectures such as ResNet \cite{ResNet} and DenseNet \cite{DenseNet}. Different than these previous neural network design problems, we propose to utilize such connections in an unrolled RNN architecture for solving inverse problems, which readily comes with its domain-specific design considerations. We compare the conventional unrolling and the proposed history-cognizant unrolling of the PGD algorithm in accelerated multi-coil MRI reconstruction using the fastMRI knee dataset \cite{fastmriRadiologyAI}, showing both visual and quantitative improvement in reconstruction quality with the proposed methodology. We also extend the same unrolling strategy to other optimization algorithms and investigate the performance improvement.

\section{Theory}
In this paper, boldface lowercase symbols denote vectors, boldface uppercase symbols denote matrices and $(\cdot)^H$ represents the matrix conjugate transpose operation.

\begin{figure}[!t]
    \begin{center}
            \includegraphics[width=\columnwidth]{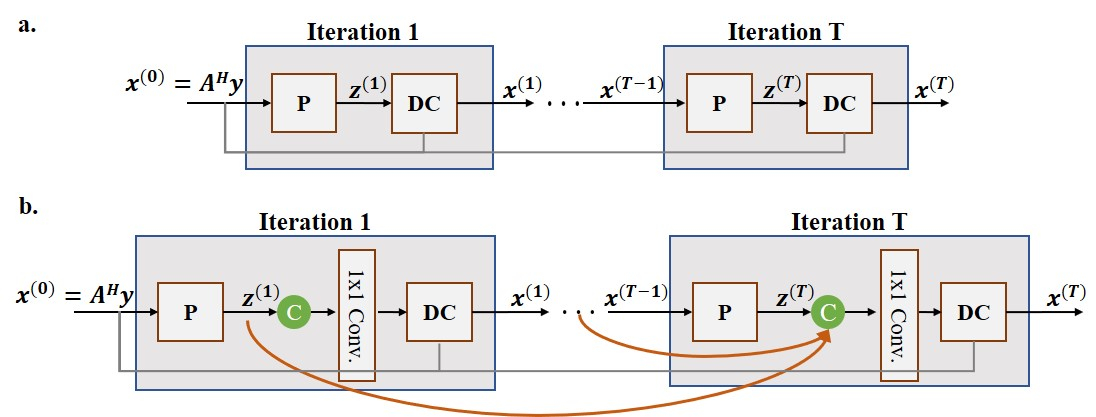}
    \end{center}
     \vspace{-.3cm}
    \caption 
    {\label{fig:netarchi} (a) The conventional unrolling for iterative algorithms for inverse problems and (b) the proposed history-cognizant unrolling leading to a dense recurrent neural network architecture. In both cases, the algorithms are unrolled for $T$ iterations, where each unrolled iteration consists of proximal (or regularization) unit (P), and a data consistency unit (DC), with $\mathbf{x}^{(0)}=\mathbf{A}^H\mathbf{y}$ as input to the network. The P unit is implemented implicitly via a neural network, while the DC unit corresponds to a gradient descent step for the proximal gradient descent algorithms and typically to conjugate gradient for variable splitting methods. For the proposed history-cognizant unrolling, an additional $1\times1$ convolutional layer combines all previous proximal operator outputs prior to the DC unit, where C denotes concatenation.}
    \vspace{-.25cm}
\end{figure} 

\subsection{Forward Model and Inverse Problem}
In a multi-coil MRI system with $n_c$ coils, let $\mathbf{y}$ be the undersampled noisy data from all coils and ${\bf x}$ be the underlying image to be recovered. The forward linear operator in this system, $\mathbf{A}: \mathbb{C}^{M \times N} \to \mathbb{C}^P$, includes sensitives of the receiver coil array and a partial Fourier matrix for undersampling in k-space, i.e. spatial Fourier domain \cite{pruessmann2001advances}. Thus, the forward model is given as 
\begin{equation}
\label{equ:forwardmodel}
\mathbf{y} = \mathbf{A}\mathbf{x} + \mathbf{n},   
\end{equation}
where $\mathbf{n}$ is measurement noise. The corresponding inverse problem 
is typically ill-conditioned. Hence, regularization is frequently used, leading to a regularized least-squares method:
\begin{equation}\label{equ:regularizedls}
\arg \min_{\mathbf{x}} \|\mathbf{y}-\mathbf{A}\mathbf{x}\|^2_2 + \mathcal{R}(\mathbf{x}),
\end{equation}
where the first term enforces consistency with measurement data and $\mathcal{R}(\cdot)$ denotes a regularizer. In classical reconstruction approaches, regularizers have included Tikhonov \cite{pruessmann2001advances, sutton2003fast}, edge-preserving functions \cite{fessler2020optimization} or sparsity-promoting terms in pre-defined domains \cite{lustig2007sparse, akccakaya2010compressed,block2007undersampled, knoll2011second}. A variety of algorithms can iteratively solve the optimization problem in (\ref{equ:regularizedls}) such as PGD \cite{combettes2011proximal}, VS \cite{goldstein2009split, ramani2010parallel,  boyd2011distributed} and PD \cite{chambolle2011first} methods.

\subsection{Physics-Based DL Reconstruction by Unrolling PGD}
Several physics-based DL methods have unrolled a PGD-based method \cite{mardani2017recurrent, IstaNet, kellman2019physics, schlemper2017deep} for a fixed number of iterations. Such PGD-based unrolling alternates between a proximal operation that is implicitly defined by a learned neural network, and a gradient descent step to enforce data consistency, as follows:
\begin{subequations}
\begin{align}
\mathbf{z}^{(i)} &=  \text{Prox}_{\mathcal{R}}(\mathbf{x}^{(i-1)})=\notag \\
&=  \arg \min_{\bf z} \|\mathbf{x}^{(i-1)}-\mathbf{z}\|_{2}^2 +\mathcal{R}(\mathbf{z}) \label{equ:pgd-updates-prox}\\
\mathbf{x}^{(i)} 
&=\mathbf{z}^{(i)}+\mu_i\mathbf{A}^H(\mathbf{y}-\mathbf{A}\mathbf{z}^{(i)})
\label{equ:pgd-updates-dc}
\end{align}
\end{subequations}
where $\mu_i$ is the gradient descent step size at unrolled iteration $i$, and the proximal operation in (\ref{equ:pgd-updates-prox}) is performed by a neural network. This leads to the schematic in \textbf{Figure \ref{fig:netarchi}a}, where the unrolled RNN consists of  multiple blocks of proximal (P) and data consistency (DC) units. The former is a neural network and the latter implements a gradient descent step based on the known forward encoding operator. The unrolled RNN is then trained end-to-end to learn the mapping from the acquired sensor-domain data to reference images for medical imaging reconstruction.  

\subsection{Proposed History-Cognizant PGD Unrolling with Dense Recurrent Neural Networks}
One of the major appeals of the PGD unrolling scheme is its computational simplicity. However, PGD algorithm itself has a slow convergence rate \cite{fessler2020optimization}, potentially requiring a large number of iterations (i.e., a large number of blocks in \textbf{Figure \ref{fig:netarchi}a}) when unrolled into a neural network for sufficient reconstruction quality. In turn, this increases memory requirements and impedes training complexity. On the other hand, several methods have been proposed to accelerate convergence rate of PGD \cite{nesterov2005smooth, nesterov2013gradient, nesterov2019gradient, beck2009fast,  Chambolle15, hong13complexity, tseng08acc, becker2011nesta}. The underlying theme for these methods is to calculate the gradient step not only at the most recent proximal operation output, but at a linear combination of all past such outputs. 

We propose to adapt this general accelerated PGD approach for network unrolling. In this setting, the accelerated PGD methods involve the following steps:
\begin{subequations}
\begin{align}
& \mathbf{z}^{(i)} = \text{Prox}_{\mathcal{R}}(\mathbf{x}^{(i-1)})\label{equ:hc-pgd-updates-prox}\\
&\mathbf{v}^{(i)} = \mathbf{F}(\mathbf{z}^{(i)}, \cdots, \mathbf{z}^{(1)})\\
& \mathbf{x}^{(i)}=\mathbf{v}^{(i)}+\mu_i\mathbf{A}^H(\mathbf{y}-\mathbf{A}\mathbf{v}^{(i)})\label{equ:hc-pgd-updates-dc},
\end{align}
\end{subequations}
where $\mathbf{F}(\cdot)$ forms a linear combination of the previous proximal operator outputs. As a special case, Nesterov's method uses the previous two outputs \cite{nesterov2013gradient}, while more general versions have also been studied, in which all the past iterations are used \cite{nesterov2019gradient, tseng08acc, becker2011nesta}. 

For our unrolled network, we propose to learn $\mathbf{F}(\cdot)$ during training instead of specifying a pre-determined form for the linear combination of previous iterates $\mathbf{F}(\cdot)$ as in Nesterov's method \cite{nesterov2013gradient}. This history-cognizant unrolling of the accelerated PGD algorithms leads to a dense recurrent neural network architecture (Dense-RNN) with skip connections across all unrolled iterations, as depicted in \textbf{Figure \ref{fig:netarchi}b}. In addition to the P and DC units to perform proximal operation (or regularization) and data consistency updates, each unrolled iteration also consists of a $1\times1$ convolutional layer implementing $\mathbf{F}(\cdot)$ on the concatenation (denoted by C) of all past proximal operator outputs prior to the DC unit. We refer to this method as history-cognizant PGD (HC-PGD) in the rest of this study. 

\section{Multi-Coil MRI Reconstruction Experiments and Results}
The proposed history-cognizant algorithm unrolling approach was compared to the conventional algorithm unrolling in multi-coil MRI reconstruction. The input $\mathbf{x}_0$ to the unrolled network in this system is the zero-filled SENSE-1 \cite{sotiropoulos2013effects} image, which is given as $\mathbf{A}^H\mathbf{y}$ for ${\bf A}$ including normalized coil sensitivities.

\begin{figure}[!t]
    \begin{center}
          \includegraphics[width=\columnwidth]{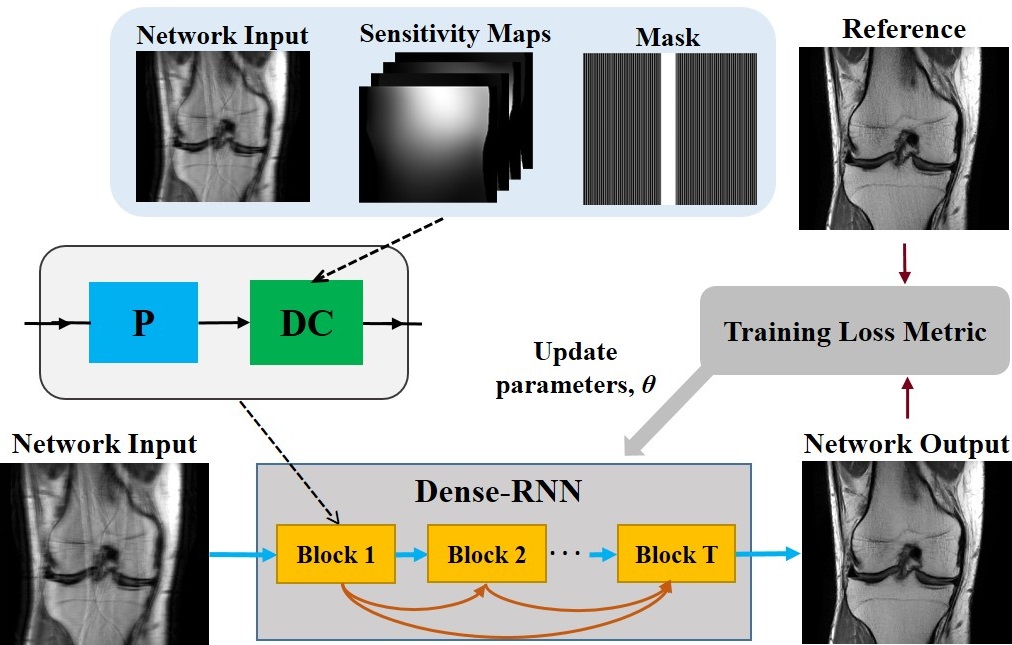}
    \end{center}
    \vspace{-.3cm}
    \caption 
    {\label{fig:overview} Summary of the training procedure in the proposed history-cognizant unrolling of inverse problem optimization algorithms with a Dense-RNN architecture for accelerated DL MRI reconstruction. Parameters of the unrolled network, which are dominantly characterized by proximal (P) operator units, are updated in each epoch based on a given loss metric using conventional end-to-end training. Data consistency (DC) units utilize the acquired data, coil sensitivity maps and undersampling mask in updates. The uniform undersampling mask utilized for training the network is also depicted.}
    \vspace{-.25cm}
\end{figure}

\subsection{Knee MRI Datasets} \label{sec:3a}
Knee MRI datasets were obtained from the New York University (NYU) fastMRI initiative database \cite{fastmriRadiologyAI} to investigate accelerated MRI reconstruction using conventional unrolled RNN and Dense-RNN architectures. Data were acquired without any acceleration on a clinical $3$T system (Magnetom Skyra, Siemens, Germany) with a 15-channel knee coil. Data acquisition was performed using 2D turbo spin-echo sequences in coronal orientation with proton-density (Coronal PD) and proton-density weighted with fat suppression (Coronal PD-FS) weightings. Relevant imaging parameters were: resolution = $0.49 \times 0.44$mm$^2$, slice thickness = $3$mm , matrix size = $320\times368$ for both datasets.

All data were uniformly sub-sampled at an acceleration rate of $4$ by keeping the central $24$ lines using the mask provided in the fastMRI database \cite{fastmriRadiologyAI} retrospectively. Uniform undersampling for 2D acquisitions is typically considered a more challenging setup for DL reconstruction than random undersampling \cite{hammernik2018learning}. A $24\times24$ central window was used to generate coil sensitivity maps using ESPIRiT \cite{uecker2014espirit}. 
The training sets consisted of $300$ slices from $15$ subjects for each sequence. Testing was performed on $10$ different subjects for each sequence. 

\subsection{Implementation Details} \label{sec:3b}
The optimization algorithms were unrolled for $T = 10$ iterations. The proposed HC-PGD algorithm was implemented using the Dense-RNN architecture in \textbf{Figure \ref{fig:netarchi}b}. Proximal operator unit outputs are first concatenated and then combined using convolutional layers with a $1\times1$ kernel size and $2$ output channels (real and imaginary components) before being inputted to the subsequent data consistency unit. On the other hand, the PGD algorithm was implemented using a conventional architecture that only includes P and DC units, without skip connections, concatenation of previous iterates or $1 \times 1$ convolutions.

The same CNN was used to implement the proximal operation for both PGD and HC-PGD. A residual network (ResNet) that has been established in a different regression problem was utilized \cite{timofte2017ntire}. This CNN consisted of $15$ sub-residual blocks each having two convolutional layers of kernel size = $3\times3$ and output channels = $64$. The first and second convolutional layers in each sub-residual block are followed by a ReLU activation function and a scaling factor of $0.1$, respectively. In addition to sub-residual blocks, two input and output convolutional layers without any nonlinear activation function are to match number of desired channels. For both algorithm unrolling schemes, the proximal units shared parameters across unrolled iterations \cite{aggarwal2018modl}, leading to a total of 592,138 and 592,358 parameters for PGD and HC-PGD, respectively. In addition to network parameters, the gradient descent step sizes ($\mu_i$) were also learned during training.

All networks were trained end-to-end, as summarized in \textbf{Figure \ref{fig:overview}}, by using Adam optimizer to minimize 
\begin{equation}
    \min_{\bm \theta} \frac1N \sum_{i=1}^{N} \mathcal{L}\Big({\bf x}^i_{ref}, \: f({\bf y}^i, {\bf A}^i; {\bm \theta}) \Big),
\end{equation}
where ${\bf x}^i_{ref}$,  ${\bf y}^i$ and ${\bf A}^i$ are the reference SENSE-1 image, undersampled data and forward encoding operator for slice $i$, respectively and $N$ is the number of training slices. $f({\bf y}^i, {\bf A}^i; {\bm \theta})$ denotes the network output for slice $i$ with the corresponding measurements and forward encoding operator, while ${\bm \theta}$ contains the trainable parameters of the networks. A normalized $\ell_1-\ell_2$ loss \cite{knoll2019deep} was used for $\mathcal{L}(\cdot, \cdot)$:
\begin{equation}
    {\cal L}({\bf u_1}, {\bf u_2}) = \frac{||{\bf u_1} - {\bf u_2}||_2}{||{\bf u_1}||_2} + \frac{||{\bf u_1} - {\bf u_2}||_1}{||{\bf u_1}||_1},
\end{equation}
with $\mathbf{u}_1$ and $\mathbf{u}_2$ representing reference fully-sampled and network output images, respectively. Normalization of the $\ell_1$ and $\ell_2$ loss terms facilitates adjusting the ratio of contribution from each term which is equal after normalization in this case \cite{knoll2019deep}. Training was performed over 100 epochs. All algorithms were implemented using TensorFlow v1.14 in Python v3.6, and processed on a workstation with an Intel E5-2640V3 CPU ($2.6$GHz and $256$GB memory), and two NVIDIA Tesla V100 GPUs with $32$GB memory.

\subsection{Performance of HC-PGD versus PGD Unrolling}
Experiments were conducted on both Coronal PD and PD-FS datasets to compare the reconstruction quality of the conventional PGD and the proposed HC-PGD unrolling with the Dense-RNN architecture. A total of 380 slices from 10 subjects (excluded from training) were utilized for testing. Reconstructed images were evaluated quantitatively using peak signal to noise ratio (PSNR) and structural similarity index (SSIM) metrics with respect to fully-sampled reference image.

\begin{figure}[!h]
    \begin{center}
            \includegraphics[trim={0 0 0 0.05in},clip, width=\columnwidth]{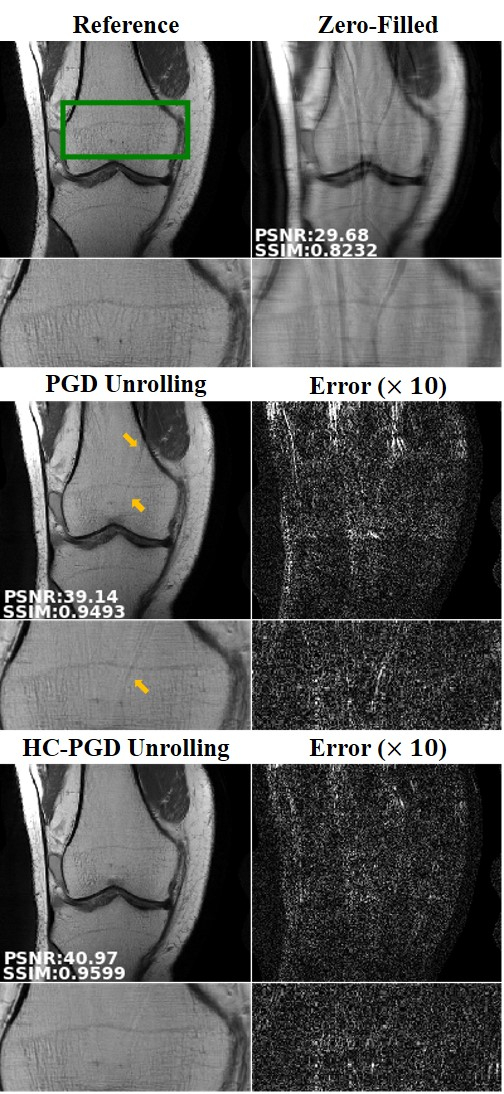}
    \end{center}
    \vspace{-.3cm}
    \caption 
    {\label{fig:coronalpd1} A representative slice from the Coronal PD dataset with 4-fold acceleration rate using a uniform undersampling pattern, reconstructed with DL-MRI reconstruction based on the conventional PGD and the proposed HC-PGD unrolling, as well as error images with respect to reference fully-sampled image scaled by a factor of 10. Zero-filled image corresponds to the input to the network, ${\bf x}^{(0)}.$ The green rectangle in the reference image marks the zoomed area shown in the second row for each method. Residual aliasing artifacts, visible in PGD unrolling results, have been suppressed using the HC-PGD scheme. Improvements in the quantitative metrics, shown on the lower-left corner of the images, also align with these visual observations.}
    \vspace{-.3cm}
\end{figure} 

\textbf{Figure \ref{fig:coronalpd1}} depicts a representative slice from the Coronal PD dataset reconstructed using DL-MRI with PGD and HC-PGD unrolling methods. The first row shows the reference fully-sampled image, as well as the zero-filled acquired image, which serves as the input ${\bf x}^{(0)}$ to the network. 
Difference images with respect to the reference image, scaled by a factor of 10, are also provided to facilitate comparison by showing errors. DL-MRI based on the proposed HC-PGD unrolling was able to suppress the strong aliasing artifacts successfully, while residual aliasing is visible on the DL-MRI reconstruction with conventional PGD unrolling. For all methods, there is a slight degree of blurring in final images, consistent with previous reports on DL-MRI methods with similar training loss functions \cite{hammernik2018learning}. Quantitative evaluations, shown in the lower left corner of the images, confirm the improvement using the proposed history-cognizant unrolling with Dense-RNN architecture compared to conventional algorithm unrolling.

\begin{figure} [!t]
    \begin{center}
          \includegraphics[trim={0 0 0 0.05in},clip, width=\columnwidth]{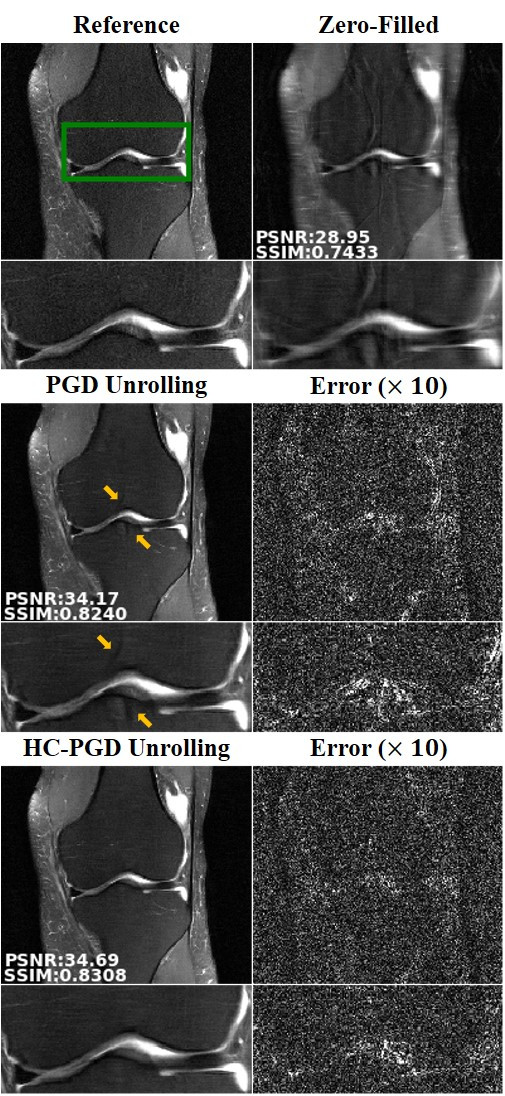}
    \end{center}
    \vspace{-.3cm}
    \caption 
    {\label{fig:coronalpdfs1} A slice from the Coronal PD-FS dataset with 4-fold acceleration using uniform undersampling, reconstructed with DL-MRI based on conventional PGD and proposed HC-PGD unrolling methods, and error images (scaled by 10). The green rectangle in the reference image mark the zoom area. The proposed HC-PGD unrolling successfully removes some residual artifacts that are visible in the PGD, and improves the quantitative metrics.}
    \vspace{-.3cm}
\end{figure} 
The reconstructed images for a representative slice from the coronal PD-FS dataset, the difference images scaled by a factor of 10, as well as reference and zero-filled images are shown in \textbf{Figure \ref{fig:coronalpdfs1}}. There are slight residual aliasing artifacts in the DL-MRI reconstruction with PGD unrolling, which are removed further by the proposed HC-PGD unrolling. The improvement in quantitative metrics also align with these observations.

\begin{figure} [t]
    \begin{center}
         \includegraphics[width=1\columnwidth]{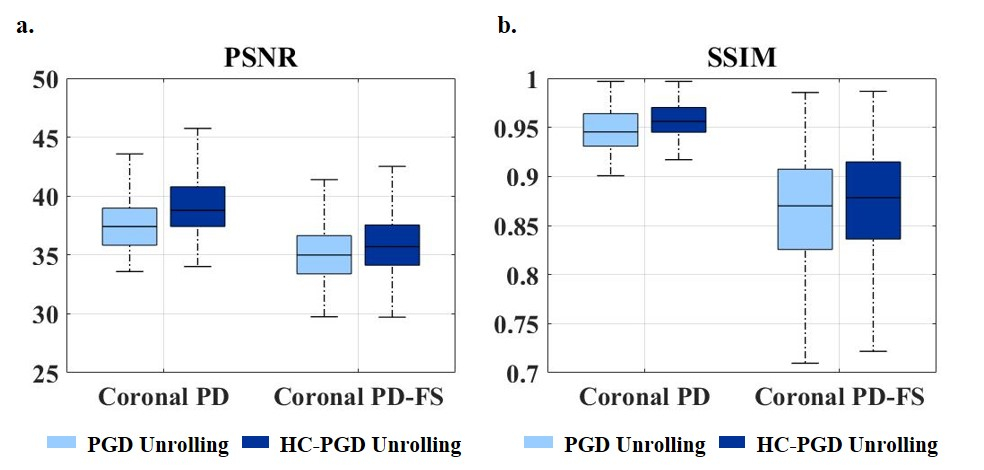}
    \end{center}
    \vspace{-.3cm}
    \caption 
    {\label{fig:boxplots} Summary of the quantitative metrics for reconstruction quality, (a) PSNR and (b) SSIM in Coronal PD and Coronal PD-FS datasets using the conventional PGD and the proposed HC-PGD methods. The boxes mark the interquartile range and median of the metrics. The proposed history-cognizant unrolling of the PGD has visibly enhanced reconstruction, based on both metrics and in both datasets. Statistical analysis based on Wilcoxon signed rank test further confirm that all improvements are statistically significant.}
    \vspace{-.3cm}
\end{figure} 

\textbf{Figure \ref{fig:boxplots}} summarizes the quantitative analysis of the DL-MRI reconstructions using PSNR and SSIM metrics for the PGD and HC-PGD unrollings for both the coronal PD and PD-FS datasets. The box plots depict the interquartile range with the center at the median of the metric for the corresponding reconstruction method. Both metrics are improved over the whole test sets when using the proposed history-cognizant algorithm unrolling over the conventional unrolling approach. Further statistical tests were performed using the Wilcoxon signed rank test with a significance level of $P<0.05$, indicating that all differences are statistically significant based on both metrics and in both datasets. 

\subsection{Insensitivity of Dense-RNN Improvement to the Regularization CNN Choice}
\begin{figure*} [t]
    \begin{center}
         \includegraphics[width=\textwidth]{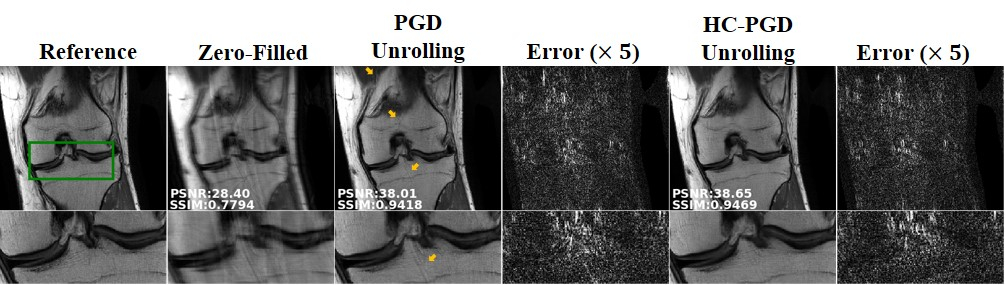}
    \end{center}
    \vspace{-.3cm}
    \caption 
    {\label{fig:unet} Results of changing the proximal operator CNN, with a slice from the Coronal PD dataset with 4-fold uniform undersampling, reconstructed with DL-MRI reconstruction using the PGD and HC-PGD methods unrolled in a network with a U-Net design for the proximal operator units, along with Error images ($\times 5$), with the green rectangle marking the zoom-up area. Similar to the ResNet experiments, the proposed HC-PGD outperforms PGD in suppressing visible residual artifacts.}
    \vspace{-.25cm}
\end{figure*} 

In order to verify that the improvements between PGD and HC-PGD unrolling are not limited to the choice of CNN used for the proximal operation, the two approaches in \textbf{Figure \ref{fig:netarchi}} were also implemented using a U-Net architecture for the regularization CNN. This was based on the U-net used in \cite{fastmriRadiologyAI}, which includes contracting and expanding paths up to $256$ channels without batch normalization. 
\textbf{Figure \ref{fig:unet}} depicts the results of this experiment on the coronal PD datasets, where the proposed history-cognizant unrolling outperforms the conventional unrolling methodology both visually and quantitatively. This highlights that the improvement from Dense-RNN is not limited to a specific choice of network architecture for the proximal operation.

\subsection{Performance Evaluation for Random Undersampling}
\label{sec:3d}
\begin{figure*}[!b]
    \begin{center}
          \includegraphics[width=1\textwidth]{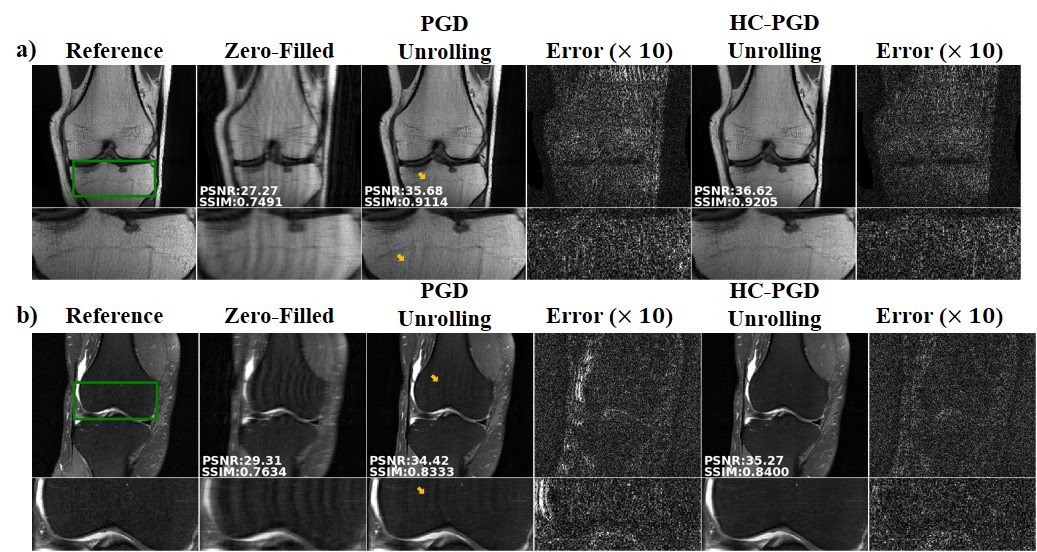}
    \end{center}
     \vspace{-.3cm}
    \caption 
    {\label{fig:random} Results on random undersampling experiments from (a) the Coronal PD and (b) Coronal PD-FS datasets, at a $4$-fold acceleration rate, reconstructed with DL-MRI reconstruction based on the conventional PGD and the proposed HC-PGD unrolling, as well as error images ($\times 10$), with the green rectangle showing the zoom-up area. Regardless of the undersampling pattern, the history-cognizant unrolling approach, HC-PGD suppresses the residual aliasing visible in the conventional PGD unrolling, as marked by the arrows. The quantitative metrics align with these observations.}
    \vspace{-.25cm}
\end{figure*} 
In order to evaluate performance of the proposed history-cognizant unrolling approach in random undersampling, we utilized an observation a previous study \cite{hammernik2018learning, knoll2019assessment}, showing that physics-based networks trained on uniformly undersampled data generalizes to random undersampling patterns at the same rate, even though the opposite is not necessarily true. Thus, we  employed the networks trained on uniformly undersampled data, as detailed in Section \ref{sec:3b}, to reconstruct test slices which were randomly undersampled at $4$-fold acceleration. The corresponding results, shown in \textbf{Figure \ref{fig:random}} indicate that the improvement trend remains similar to the uniform sampling scenario with the proposed HC-PGD method outperforming the conventional PGD in suppressing residual artifacts in both Coronal PD and PD-FS datasets, and improving the quantitative evaluation metrics.

\subsection{Extensions to Other Optimization Algorithms}\label{sec:ext-to-vs}
While PGD unrolling is commonly used for DL-MRI reconstruction \cite{mardani2017recurrent, IstaNet, kellman2019physics, schlemper2017deep}, alternative algorithms have also been utilized \cite{yang2018admm, deepADMMnet, ye2018deep, aggarwal2018modl, qin2018convolutional, duan2019vs, yaman2019self, yaman2019selfMRM}. One such approach relies on VS methods \cite{yang2018admm, deepADMMnet, ye2018deep, aggarwal2018modl, qin2018convolutional, duan2019vs, yaman2019self, yaman2019selfMRM} for solving the objective function in Equation (\ref{equ:regularizedls}), which decouples data consistency and regularization terms into two blocks of variables \cite{fessler2020optimization} by introducing auxiliary variables to formulate a constrained objective function. This constrained problem is then relaxed to an unconstrained problem by enforcing similarity between these variables either via a quadratic penalty \cite{ye2018deep, aggarwal2018modl, qin2018convolutional, duan2019vs, yaman2019self, yaman2019selfMRM} or via an augmented Lagrangian approach \cite{yang2018admm, deepADMMnet}.

In the case of variable splitting with quadratic penalty (VSQP), the following objective function is used:
\begin{equation}\label{equ:regularized}
\arg \min_{\mathbf{x, z}} \|\mathbf{y}-\mathbf{A}\mathbf{x}\|^2_2 + \mathcal{R}(\mathbf{z}) + \beta \|\mathbf{x}-\mathbf{z}\|^2_2,
\end{equation}
where $\mathbf{z}$ is an auxiliary variable and $\beta$ is the quadratic penalty parameter. When this algorithm is unrolled as a neural network \cite{ye2018deep, aggarwal2018modl, qin2018convolutional, duan2019vs, yaman2019self, yaman2019selfMRM}, reconstruction alternates between a proximal operation that is again learned by a neural network and data consistency, similar to PGD unrolling, as follows:
\begin{subequations}
\begin{align}
\mathbf{z}^{(i)} = &  \text{Prox}_{\mathcal{R}}(\mathbf{x}^{(i-1)}) \label{equ:vs-updates-reg}
\\
\mathbf{x}^{(i)} = & (\mathbf{A}^H\mathbf{A}+\mu \mathbf{I})^{-1}(\mathbf{A}^H\mathbf{y} + \beta\mathbf{z}^{(i)}) \label{equ:vs-updates-dc}
\end{align}
\end{subequations}
where regularization update in (\ref{equ:vs-updates-reg}) is performed by a neural network and $\beta$ was absorbed into the regularizer for notational convenience. The data consistency update in (\ref{equ:vs-updates-dc}) is computationally challenging in many practical scenarios, such as multi-coil MRI reconstruction. Thus, this data consistency update can be solved iteratively via conjugate gradient (CG) to avoid matrix inversion \cite{aggarwal2018modl}.

\begin{figure*}[!t] \label{fig:admm-pd}
    \begin{center}
          \includegraphics[trim={0 0 0 0},clip, width=\textwidth]{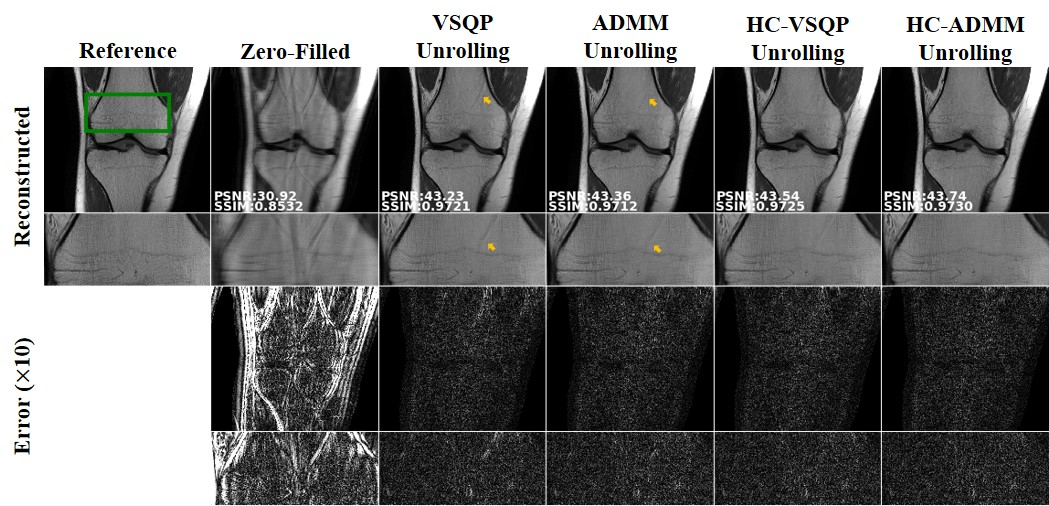}
    \end{center}
     \vspace{-.3cm}
    \caption{
    \label{fig:admm-pd}Example slice from the Coronal PD with $4$-fold uniform undersampling using DL-MRI reconstruction based on the conventional VSQP and ADMM, and the proposed HC-VSQP and HC-ADMM unrolling, as well as error images ($\times 10$), with the green rectangle showing the zoomed area. The history-cognizant unrolling approaches, HC-VSQP and HC-ADMM suppress the residual aliasing visible in the conventional VSQP and ADMM unrolling, marked by the arrows. Improvements are also observed in the quantitative metrics.}
    \vspace{-.3cm}
\end{figure*} 

\begin{figure*}[!b]
    \begin{center}
          \includegraphics[trim={0 0 0 0},clip, width=\textwidth]{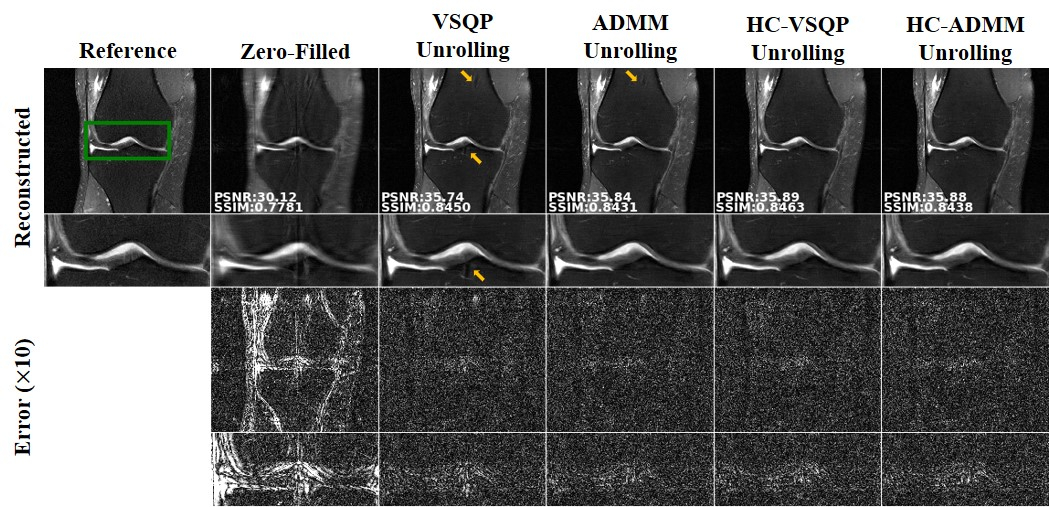}
    \end{center}
     \vspace{-.3cm}
    \caption 
    {\label{fig:admm-pdfs} Example slice from the Coronal PD-FS with $4$-fold uniform undersampling using DL-MRI reconstruction based on the conventional VSQP and ADMM, and the proposed HC-VSQP and HC-ADMM unrolling, as well as error images ($\times 10$), with green rectangle showing the zoomed area. The reconstructions from the history-cognizant unrolling approach, HC-VSQP and HC-ADMM suppress the residual aliasing visible in the conventional VSQP and ADMM unrolling, marked by the arrows. }
    \vspace{-.3cm}
\end{figure*}

\begin{figure*}[!t] \label{fig:nesterov-pd}
    \begin{center}
          \includegraphics[trim={0 0 0 0},clip, width=0.88\textwidth]{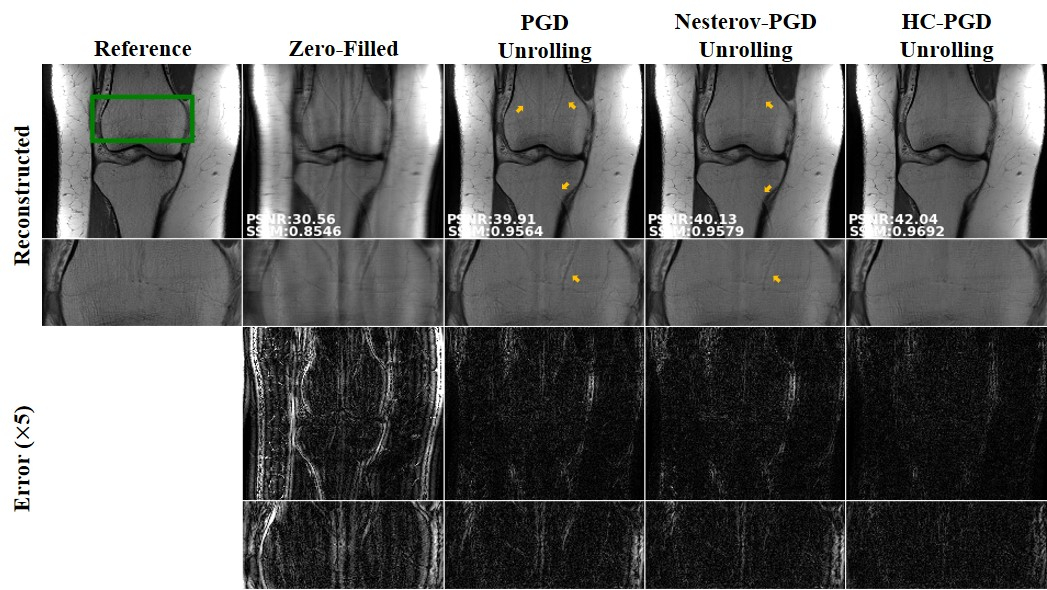}
    \end{center}
     \vspace{-.3cm}
    \caption{
    \label{fig:nesterov-pd} The effect of Nesterov-type unrolling versus full history cognizant unrolling in HC-PGD for a slice from the Coronal PD dataset with $4$-fold uniform undersampling, along with Error images ($\times5$), with the green rectangle marking the zoom-up area. Aliasing artifacts remain in the reconstructed images from conventional PGD and Nesterov-type unrolling, while these are suppressed with the proposed HC-PGD unrolling.}
    \vspace{-.3cm}
\end{figure*} 

\begin{figure*}[!b] \label{fig:conv-pd}
    \begin{center}
          \includegraphics[trim={0 0 0 0},clip, width=0.88\textwidth]{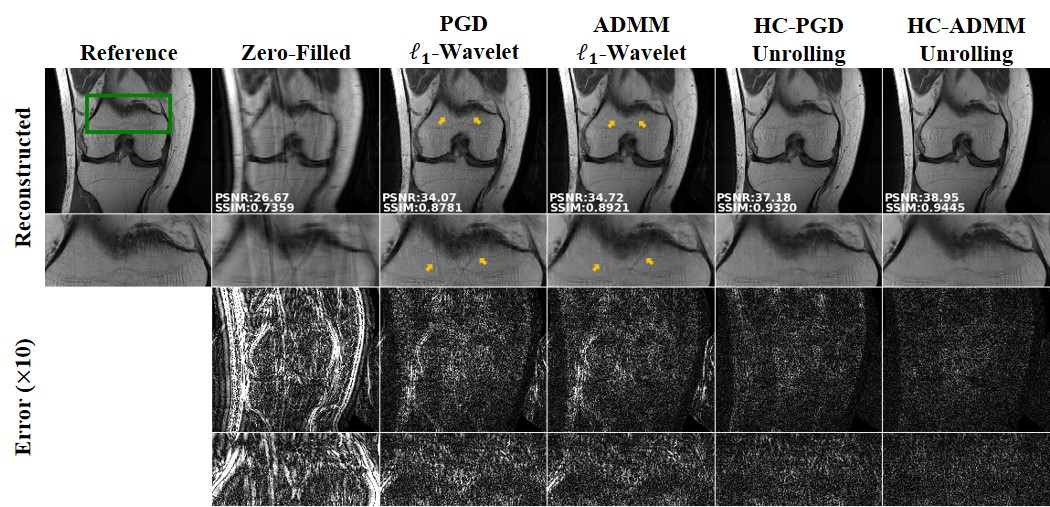}
    \end{center}
     \vspace{-.3cm}
    \caption{
    \label{fig:conv-pd} Comparison of traditional non-DL PGD and ADMM, both using $\ell_1$-wavelet regularization, against DL methods based on HC-PGD and HC-ADMM unrolling for a Coronal PD slice with 4-fold uniform undersampling, along with error images ($\times10$), with the green rectangle marking the zoom-up area. Aliasing artifacts remain in the reconstructed images from traditional PGD and ADMM, while DL versions of these algorithms successfully suppress these artifacts, resulting in superior reconstruction quality and quantitative metrics, consistent with previous studies \cite{hammernik2018learning}.}
    \vspace{-.3cm}
\end{figure*}

Another line of work utilizes an augmented Lagrangian relaxation, leading to the unrolling of the ADMM algorithm \cite{yang2018admm, deepADMMnet}. The augmented Lagrangian approach used for relaxation in ADMM crucially leads to an extra intermediate step to optimize the Lagrangian parameters in addition to the regularization and data consistency updates in VSQP unrolling. This intermediate step in general improves the convergence rate of VS methods, but also leads to more memory usage which can limit the number of unrolls that can be performed with limited GPU resources.

As an extension to the HC-PGD approach, both VSQP and ADMM methods may be unrolled in a history-cognizant manner using the proposed Dense-RNN in \textbf{Figure \ref{fig:netarchi}}, building on the optimization approaches in \cite{Chambolle15, hong13complexity}. These will be referred to as the history-cognizant VSQP (HC-VSQP) unrolling and the history-cognizant ADMM (HC-ADMM) unrolling respectively. We note that the main difference between HC-PGD and HC-VSQP unrolling is that the DC units in HC-VSQP are implemented with an unrolled CG algorithm for a different least squares problem instead of a gradient step. HC-ADMM further includes the Lagrangian parameter update as in ADMM-Net \cite{yang2018admm, deepADMMnet}.

To test the applicability of the history-cognizant unrolling scheme to different baseline algorithms, DL-MRI reconstruction based on VSQP, HC-VSQP, ADMM and HC-ADMM unrollings were also implemented on the coronal PD and PD-FS datasets, using the setup described in Section \ref{sec:3b}. In each data consistency unit, CG was implemented with $10$ internal iterations. Furthermore, due to the higher memory usage of the Lagrangian update in ADMM and HC-ADMM unrollings, these were unrolled for 8 iterations due to GPU memory limitations.

\textbf{Figure \ref{fig:admm-pd}} and \textbf{Figure \ref{fig:admm-pdfs}} depict representative slices from the Coronal PD and Coronal PD-FS datasets, respectively, reconstructed using DL-MRI with VSQP, ADMM, HC-VSQP and HC-ADMM unrolling methods, along with the error images scaled by a factor of 10, as well as the reference and zero-filled images. The reconstruction from the proposed history-cognizant unrolling approaches, i.e. HC-VSQP and HC-ADMM, successfully mitigate the residual artifacts apparent in the conventional VSQP and ADMM results, shown by the yellow arrows. Improvements are also observed in the corresponding quantitative metrics.
\begin{table*}[!t] 
\begin{adjustbox}{width=0.9\textwidth,center}
\begin{tabular}{|c|c|c|c|c|c|c|} 
\hline
              & \begin{tabular}[c]{@{}c@{}}\\\vspace{-2mm}PGD \\\\\end{tabular} & \begin{tabular}[c]{@{}c@{}}\\\vspace{-2mm}HC-PGD\\\\ \end{tabular} & \begin{tabular}[c]{@{}c@{}}\\\vspace{-2mm}VSQP\\\\\end{tabular} & \begin{tabular}[c]{@{}c@{}}\\\vspace{-2mm}HC-VSQP\\\\\end{tabular}&
              \begin{tabular}[c]{@{}c@{}}\\\vspace{-2mm}ADMM\\\\\end{tabular} & \begin{tabular}[c]{@{}c@{}}\\\vspace{-2mm}HC-ADMM\\\\\end{tabular}\\ \hline
Coronal PD   & $197\pm4.2$ ms                                                   & $197\pm7.2$  ms                                                     & $340\pm6.8$   ms                                               & $325\pm6.1$  ms        & $316\pm5.0$   ms         & $320\pm6.0$  ms                                  \\ \hline
Coronal PD-FS & $197\pm5.6$  ms                                                 & $198\pm6.6$  ms                                                   & $339\pm5.6$     ms                                             & $325\pm6.2$ ms             & $321\pm6.1$   ms         & $318\pm6.1$  ms          \\ \hline
\end{tabular}
\end{adjustbox}
\caption{Average and standard deviation (in ms) of reconstruction time over all test slices for PGD, HC-PGD, VSQP, HC-VSQP, ADMM and HC-ADMM in Coronal PD and Coronal PD-FS datasets. History cognizant unrolling has very slight impact on reconstruction time compared to its conventional counterpart. Furthermore, VSQP and ADMM methods with CG-based data consistency units, are considerably slower than PGD methods.}
\label{tbl:testingtime}
\end{table*}

\section{Discussion}
In this study, we proposed a history-cognizant method for unrolling optimization problems to solve regularized inverse problems in accelerated MRI using physics-based DL reconstruction.  The proposed unrolling method builds on optimization approaches that utilize a history of the previous iterates via their linear combinations in updating the current estimate \cite{nesterov2005smooth, nesterov2013gradient, nesterov2019gradient, beck2009fast, Chambolle15, hong13complexity}. While these optimization procedures have been traditionally proposed to improve convergence rates or provide stronger theoretical guarantees, in the context of algorithm unrolling for physics-based DL reconstruction, they are used to enhance reconstruction quality without incurring additional computational complexity. Our method was pre-dominantly motivated by accelerated PGD approaches \cite{nesterov2005smooth, nesterov2013gradient, nesterov2019gradient, beck2009fast,Chambolle15, hong13complexity} and was shown to improve upon conventional PGD unrolling for multi-coil MRI reconstruction for multiple scenarios. We also explored how the proposed unrolling can be applied to VS methods  for solving a regularized least squares problem motivated by results from optimization theory  \cite{Chambolle15, hong13complexity}, and showed that it can improve over the conventional unrolling approach.

Our work makes several novel contributions to domain enriched learning for medical imaging, specifically from the perspective of signal processing. Our main contribution is the adoption of analytical insights for improving convergence from optimization theory to improve algorithm unrolling for deep learning reconstruction. Specifically, these insights were used to add skip connections in the unrolled neural network in a principled manner, termed history-cognizant unrolling. In turn, this led to a dense RNN architecture in a regression problem for the first time, while maintaining interpretability due to the connections to optimization theory. This kind of \emph{principled approach for neural network design} is especially important for the broader signal processing community, since most existing neural network design principles from other communities are heuristic, without analogous counterparts based in theory. Finally, the proposed history-cognizant unrolling improved upon conventional unrolling for a number of different optimization algorithms, both quantitatively and visibly for various accelerated MRI scenarios. Importantly, this improvement was achieved without a cost in the number of trainable parameters, in training time or in computational complexity.

The proposed history-cognizant unrolling methodology leads to several skip connections across iterations, leading to a densely-connected unrolled neural network. Dense neural networks have been shown to improve training by facilitating information flow through the network in forward propagation and by relieving issues concerned with gradient vanishing in backward propagation \cite{DenseNet}. One of the advantages of these dense connections in the proposed unrolling is that the only computational overhead compared to conventional unrolling is the calculation of the $1 \times 1$ convolutions from the previous iterations. Thus, once trained, the computational time is similar to the conventional counterpart, as depicted in Table I.

One of the most commonly used accelerated PGD approaches is based on Nesterov acceleration \cite{nesterov2013gradient}. In this case, the gradient descent is calculated at a linear combination of the previous two iterates weighted by factors $\alpha$ and $1-\alpha$ respectively. Theoretically, this approach achieves the optimal convergence rate for conventional first-order optimization methods \cite{nesterov2013gradient}. As such, Nesterov-type unrolling for DL reconstruction was readily explored in \cite{kellman2019physics} for coded-illumination microscopy, where the unrolled accelerated PGD network used the history of the two previous outputs of the proximal units. Thus, we also performed a comparison between the proposed HC-PGD unrolling and Nesterov-type PGD unrolling \cite{kellman2019physics}, using the setup in Section \ref{sec:3b}. Results in \textbf{Figure \ref{fig:nesterov-pd}} show that the proposed history-cognizant Dense-RNN architecture suppresses residual artifacts that are apparent in Nesterov-PGD unrolling, while also improving the quantitative metrics.

In our experiments, the history-cognizant unrolling  visibly improved the performance of PGD-based methods compared to its conventional counterpart for all datasets. In the VS setting, especially for the coronal PD datasets, the improvement is less pronounced. Coronal PD datasets have higher SNR compared to PD-FS datasets \cite{hammernik2018learning}. In this higher SNR case, the conventional unrolling of the VS algorithm readily provides a high-quality reconstruction. Thus, additional improvement with history cognizant unrolling of VS is less apparent than in the PGD setting.

Our experiments in multi-coil MRI reconstruction pre-dominantly utilized uniform undersampling patterns. While random undersampling patterns are commonly used in  literature for training and testing DL-based reconstruction, uniform undersampling remains the most clinically used approach, and was considered to be a harder problem compared to random undersampling in \cite{hammernik2018learning, knoll2019assessment}, as detailed in Section \ref{sec:3d}. Uniform undersampling also creates a difficult problem for conventional non-DL methods that use a pre-specified regularizers, such as an $\ell_1$-wavelet regularization term. \textbf{Figure \ref{fig:conv-pd}} shows results from traditional non-DL implementations of PGD and ADMM with an $\ell_1$-wavelet regularization term. Even though the soft-thresholding parameter was manually fine-tuned for this slice for both PGD and ADMM, residual artifacts remain and quantitative metrics are noticeably worse than the DL methods. Both observations are consistent with previous studies \cite{hammernik2018learning}. Furthermore, these traditional methods require manual tuning of the hyperparameters, which vary across subjects and slices. We also note that we did not compare our methods to more advanced non-DL methods \cite{qu2012undersampled,zhan2015fast, qu2014magnetic, zhang2020image}, as our study focused on deep learning reconstruction. Furthermore, all our experiments concentrated on physics-driven DL reconstruction methods for MRI. While non-physics driven DL methods can speed up convergence and reduce the overall training time by learning the encoding matrix \cite{giryes2018tradeoffs}, this was not explored in this work which focused on algorithm unrolling for physics-based DL methods.

In this study, parameters of the proximal units were shared across unrolled iterations, consistent with the optimization literature where the regularization step remains unchanged over iterations. However, reconstruction quality may be further improved by learning different parameters, albeit at the cost of requiring more training examples and more careful training \cite{aggarwal2018modl}. In addition, the training loss was calculated on the network output and reference target data which were both in image domain. While image-domain loss is a common practice in the literature, the loss defined over multi-coil k-space data may provide additional benefits for reducing artifacts \cite{yaman2019self, yaman2019selfMRM, tamir2019unsupervised}, but this was not explored in the current study.

To the best of our knowledge, the Dense-RNN architecture that arises from the proposed history-cognizant unrolling have not been used for regression problems in general, and inverse problems in particular. However, Dense-RNNs have been independently proposed for different problems, such as scene parsing or labeling \cite{fan2019scene} where semantic information of an image segment is to be extracted. In such applications, RNNs are used for each image unit to receive the dependencies from other units through recurrent information forwarding between adjacent units, which decays for further units. In this setting, Dense-RNNs improve labeling by capturing long-range semantic dependencies among image units, highlighting a similar memory dependency as in this study.

Finally, the proposed history-cognizant unrolling combines past updates linearly. However, given the success of non-linear combination of the history of the gradients in improving the performance of gradient descent approaches \cite{kingma2014adam}, generalizing history cognizant unrolling to a non-linear scheme is an interesting extension of this work. However, this would require a theoretical underpinning for the use of non-linear combination of history gradients in solving the objective function in Equation (\ref{equ:regularizedls}), and thus is beyond the scope of this study.

\section{Conclusion}
We proposed a history-cognizant approach to unroll a given optimization algorithm for solving inverse problems in medical imaging using physics-driven deep learning. The proposed approach leads to an unrolled dense recurrent neural network architecture. From a network-design perspective, this consequent architecture enhances training and performance by facilitating information flow through the network. Accelerated MRI reconstruction experiments show that the proposed unrolling approach may considerably enhance reconstruction quality without substantially modifying computational complexity.

\section*{Acknowledgment}
MRI data were obtained from the NYU fastMRI initiative database \cite{fastmriRadiologyAI} which was acquired with the relevant institutional review board approvals as detailed in \cite{fastmriRadiologyAI}. NYU fastMRI investigators provided data but did not participate in analysis or writing of this report. A listing of NYU fastMRI investigators, subject to updates, can be found at \url{fastmri.med.nyu.edu}.

\bibliographystyle{IEEEbib}
\bibliography{reference}

\end{document}